# The adaptive filter of the yeast galactose pathway

Serge Smidtas [a], Vincent Schächter [b], François Képès [c]

[a] Genoscope-Centre National de Séquençage, CNRS UMR8030, 2 rue Gaston Crémieux, 91000, Evry, France
sergi@sergi5.com

[b] Genoscope-Centre National de Séquençage, CNRS UMR8030, 2 rue Gaston Crémieux, 91000, Evry, France
vs@genoscope.cns.fr

[c] Epigenomics Project, Genopole®, 523 Terrasses de l'Agora, 91000 Evry, France & ATGC, CNRS UMR8071/Genopole®, Evry, France
Francois.Kepes@genopole.cnrs.fr

## Abstract

In the yeast *Saccharomyces cerevisiae*, the interplay between galactose, Gal3p, Gal80p and Gal4p determines the transcriptional status of the genes required for galactose utilization. After an increase in galactose concentration, galactose molecules bind onto Gal3p. This event leads via Gal80p to the activation of Gal4p, which then induces *GAL3* and *GAL80* gene transcription. Here we propose a qualitative dynamic model, whereby these molecular interaction events represent the first two stages of a functional feedback loop that closes with the capture of activated Gal4p by newly synthesized Gal3p and Gal80p, decreasing transcriptional activation and creating again the protein complex that can bind incoming galactose molecules. Based on the differential time scales of faster protein interactions versus slower biosynthetic steps, this feedback loop functions as a derivative filter where galactose is the input step signal, and released Gal4p is the output derivative signal. One advantage of such a derivative filter is that *GAL* genes are expressed in proportion to the cellular requirement. Furthermore, this filter adaptively protects the cellular receptors from saturation by galactose, allowing cells to remain sensitive to variations in galactose concentrations rather than to absolute concentrations. Finally, this feedback loop, by allowing phosphorylation of some active Gal4p, may be essential to initiate the subsequent long-term response.

# 1. Background

Living organisms constantly adapt to fluctuations in their intra- and extra-cellular environments, in part by regulating their gene expression. Gene expression can be controlled at many levels that involve protein-DNA (transcriptional), protein-protein and protein-small molecule interactions. The process of galactose (GAL) utilization in the common yeast *Saccharomyces cerevisiae* has been thoroughly studied; yeast is known to exhibit sophisticated responses to the presence of different types of sugar in its environment. The GAL pathway is a classic example of a genetic regulatory switch, in which enzymes specifically required for the transport and catabolism of galactose are expressed only when galactose is present and repressing sugars such as glucose are absent in the cellular environment (Biggar and Crabtree, 2001).

The permease encoded by the *GAL2* gene, and possibly other hexose transporters (HXTs) transport galactose across the cell membrane. Other genes encode the enzymes required for conversion of intracellular galactose, including galactokinase (*GAL1*), uridyltransferase (*GAL7*), epimerase (*GAL10*), and phosphoglucomutase (*GAL5/PGM2*). Galactose activates the transcription of *GAL* genes from undetectable or low basal levels to high levels. The activated genes include *GAL1*, *GAL2*, *GAL3*, *GAL5*, *GAL7* and *GAL80* (Sakurai et al., 1994), but not *GAL4* (Ren et al., 2000; Ideker et al., 2001). The complex interplay of Gal4p, Gal80p, and Gal3p determines the transcriptional status of these *GAL* genes (Platt and Reece, 1998). Gal4p is a DNA-binding transcriptional activator that can bind to upstream activating sequences in the promoter regions of target *GAL* genes, thereby strongly activating their transcription. However, in the absence of galactose, Gal4p is sequestered by Gal80p and is unable to activate transcription of the *GAL* genes, although this Gal4p/80p complex appears to bind DNA (Parthun and Jaehning, 1992). The interaction between Gal4p and Gal80p is weaker in the presence of galactose (Sil et al., 1999). Gal80p and Gal3p may also form a complex, which in contrast is stabilized in the presence of galactose (Yano and Fukasawa, 1997). Gal3p overproduction, presumably by sequestering Gal80p away from Gal4p, causes galactose-independent activation of the GAL pathway (Bhat and Hopper, 1992; Peng and Hopper, 2000).

*Gal3* mutant cells are still able to activate the GAL pathway in response to galactose. However, induction requires several days rather than a few minutes in wild-type yeast, a phenomenon called long-term adaptation (LTA) (Winge and Roberts, 1948; Bhat and Murthy, 2001). It was proposed (Rohde et al., 2000) that the LTA of the GAL pathway is mediated by Gal4p phosphorylation. Indeed, when Gal4p is bound to DNA and interacts with the RNA-polymerase II holoenzyme, its serine at position 699 (S699) becomes phosphorylated by Srb10p/Cdk8p, a component of the 'Mediator' subcomplex of the holoenzyme (Hirst et al., 1999; Bhaumik and Green, 2001; Larschan and Winston, 2001). Gal4p S699 phosphorylation is necessary to amplify and maintain full *GAL* gene induction (Sadowski et al., 1996; Yano and Fukasawa, 1997; Rohde et al., 2000).

The above set of experimental observations raises two main questions. Firstly, the system responds to galactose increases rather than to absolute galactose concentration: how is this achieved (Rohde et al., 2000)? Secondly, several authors have observed that Gal4p does not become phosphorylated unless it activates transcription, yet that it is not fully active unless it is phosphorylated (Sadowski et al., 1991; Sadowski et al., 1996; Hirst et al., 1999). A satisfactory explanation for this 'chicken and egg' enigma is lacking. In this paper we propose a mathematical model of the early response to galactose and we analyze its dependence upon time delays, protein degradation rate and initial conditions. The model accounts for the above-mentioned sensitivity to galactose fluctuations. It also proposes a solution to the apparent paradox described above by showing that a feedback loop brings active Gal4p onto gene promoters, thus allowing its phosphorylation and consequent maintenance of transcriptional activation.

## 2. Qualitative Modeling of the Galactose Response

*2.1. Assumptions*

The present model deals with the early steps of galactose induction; it does not consider the events occurring after Gal4p phosphorylation. It does not emphasize the details of signal transmission from galactose to Gal4p (except in Appendix A). Thus, Gal4p appears in this model either bound to DNA, or bound to DNA and to Gal80p. Gal80p is either bound to DNA and to Gal4p, or bound to Gal3p, or unbound. An equilibrium between nuclear and cytoplasmic forms of Gal80p has been considered by other authors (Peng and Hopper, 2000; Peng and Hopper, 2002; Verma et al., 2003), but is not relevant here given the scope of our model. The order in which galactose, ATP, Gal3p and Gal80p bind together is not fully known but should have no effect on the conclusions reached with our model, which simply considers Gal80p consumption upon galactose addition.

The *GAL1* gene is a paralogue of the *GAL3* gene (Wolfe and Shields, 1997) that encodes a galactokinase, while Gal3p does not have galactokinase activity (Platt et al., 2000). Galactokinase activity is irrelevant to the present model which does not address galactose catabolism. Therefore, Gal1p and Gal3p are taken to play a similar role in GAL pathway activation, averaged over their respective abundances and inducing properties. In the model, they will be lumped together under the name of Gal3p.

*2.2. Dynamic description of the GAL system*

Figure 1 illustrates the different states of the dynamic system. Figure 2A illustrates the core regulatory mechanism. In the absence of galactose, Gal4p can bind to Gal80p and has no transcriptional activity. Following a step increase in galactose, Gal3p rapidly binds galactose, and Gal80p is consumed by being recruited in a complex with Gal3p. As the concentration of unbound Gal80p decreases, the Gal4p/80p complex is destabilized, which activates Gal4p. Activated Gal4p then initiates the slower biosynthetic reactions, transcription of the *GAL* genes including *GAL3* and *GAL80*, followed by translation into their protein products.

Following Gal4p activation, and consequent *GAL* gene expression, newly synthesized Gal3p and Gal80p shift the equilibrium back towards Gal4p inactivation. As a result, *GAL* transcriptional activity decreases back. Newly formed proteins can bind incoming galactose molecules, thus restoring sensitivity to any further galactose input. This effectively closes the feedback loop, the central point of this model.

*2.3. Model simplification*

The detailed model shown in Figure 2A is needlessly complex relative to the focus of our study: the role of the feedback loop. In this section, we show how the model could be simplified, yielding a reduced model (Fig. 2B) that features the feedback-loop and preserves the qualitative dynamics of the detailed model, while allowing deeper analysis and understanding.

The *GAL3* and *GAL80* genes are both transcriptionally regulated by Gal4p. However, the *GAL3* gene is activated about five-fold stronger than *GAL80* (Peng and Hopper, 2002). This fact may amplify or accelerate the response. Indeed, a relative increase of Gal3p with respect to Gal80p will shift complex formation towards additional Gal80p consumption, increasing further the concentration of activated Gal4p. Even though this may bring about changes in the exact response kinetics, it does not change the qualitative behavior of the system. Furthermore, concomitant overexpression of *GAL3* and *GAL80* was shown to suppress the constitutive *GAL* gene expression elicited by overexpression of *GAL3* alone (Suzuki-Fujimoto et al., 1996), suggesting that their two products play a complementary role in the reaction cascade. Accordingly, Gal3p and Gal80p are lumped together in this model as the 'Gal3/80p' complex. This simplification is relaxed in a more complex model described in Appendix A. This model may be used for simulation. However, it is not amenable to analytical understanding, as it involves six equations and eight parameters. It is shown that the results on the qualitative behavior remain unchanged (Appendix A).

In the absence of galactose, Gal3p, Gal80p and Gal4p form an inactive complex Gal4/3/80p called 'receptor' ('R'; Fig.2B). This simplifies the model, lumping a cascade of reactions into one. A 'bound receptor' ('BR') comprising Gal3p, Gal80p and galactose remains inactive or may be degraded. Finally, the total Gal4p concentration is assumed to be constant during the GAL response, as suggested by transcriptomics data (Ren et al., 2000; Ideker et al., 2001).

The three main transformations of the simplified model corresponding to the three reactions of Figure 2B are shown below (Eq. 1-3). We represent a gene, its encoded mRNA and protein as a single entity. 'G4' denotes an active Gal4p protein. The first equation pertains to the slow biosynthetic steps of transcription and translation, comprising the binding of Gal4p to the *GAL3* and *GAL80* gene promoters, and all subsequent actions until Gal3/80p molecules are newly synthesized one at a time, and Gal4p stays activated. The second equation represents the inactivation of Gal4p into its inactive form called the receptor 'R'. The third equation expresses the activation of Gal4p due to the binding of galactose ('Gal') to the receptor, yielding the bound receptor ('BR').

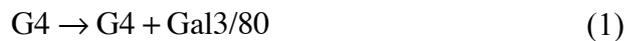
G4 → G4 + Gal3/80              (1)
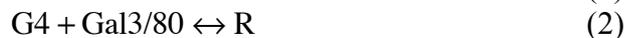
G4 + Gal3/80 ↔ R               (2)
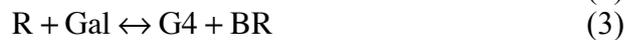
R + Gal ↔ G4 + BR              (3)

To facilitate analyses and simulations, this model can be further reduced to a two equation, two variable system readily amenable to phase plane analysis. It emphasizes the role played by the negative feedback loop in the *GAL* pathway dynamics. Combining transformations (1) and (2) into (2a), gives the following set of two equations, with three kinetic constants K.

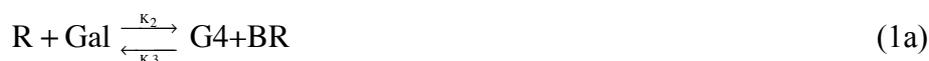
$$R + Gal \underset{K_3}{\overset{K_2}{\rightleftarrows}} G4 + BR \qquad (1a)$$

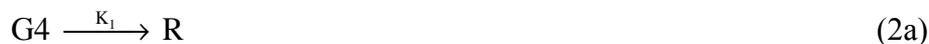
$$G4 \xrightarrow{K_1} R \qquad (2a)$$

'$S_0$' (for "Sugar") is the initial quantity of galactose, and '$R_0$' (for "Receptor") the total amount of Gal4p. These values are constant, as galactose is initially provided in a finite amount, and total Gal4p has been assumed to be a constant amount (Johnston et al., 1994; Ren et al., 2000; Ideker et al., 2001)).

$$\text{Gal} + \text{BR} = S_0 \quad (1b)$$
$$\text{R} + \text{G4} = R_0 \quad (2b)$$

This leads to two non-linear differential equations, with $K_1$ normalized to 1. From assuming a homogeneous spatial distribution and a mean cell volume of 70 μm$^3$ (Ruhela et al., 2004), the order of magnitude of concentration of galactose and proteins involved is of the order of 10$^{-6}$ M. This led to $S_0$ and $R_0$ of the order of 10$^{-6}$ M and $K_1=K_2=K_3=1$. One to 10 hours is the typical time response of the galactose switch. These values imply that the order of magnitude of time unit is 5x10$^2$s.

$$\frac{d(R)}{dt} = (K_3-K_2).\text{Gal}.R - K3.R_0.\text{Gal} - (K_3.S_0+K_1).R + R_0.(K_3.S_0+1) \quad (1c)$$

$$\frac{d(\text{Gal})}{dt} = (K_3-K_2).\text{Gal}.R + K_3.S_0.R_0 - K_3.S_0.R - K_3.R_0.\text{Gal} \quad (2c)$$

In the special case where $K_2$ is set equal to $K_3$, these equations can be simplified into linear equations:

$$\frac{d(R)}{dt} = (K_2.S_0 + 1).R_0 - (K_2.S_0 + 1).R - K_2.R_0.\text{Gal} \quad (1d)$$

$$\frac{d(\text{Gal})}{dt} = K_2.S_0.R_0 - K_2.S_0.R - K_2.R_0.\text{Gal} \quad (2d)$$

## 3. Results

### 3.1. Steady States and Phase Plane Portrait

The phase plane for equations (1c) and (2c) is the Cartesian coordinate system representing the two variables, receptor and free galactose concentrations (Fig. 3). Two nullclines are represented, that correspond to the situations where: 1) the receptor synthesis and consumption by galactose fixation are balanced, or 2) free galactose addition and galactose fixation are balanced. Thus, the nullclines correspond to steady states, where the derivatives of equations (1c) and (2c) are null (Fig. 3). The two nullclines cross each other at a stable point which corresponds to the situation where 1) and 2) simultaneously hold true. This stable point turns out to correspond to low free galactose and high receptor concentrations (Fig. 3). When $K_1$ is normalized to 1, and $K_2$ and $K_3$ are equal (equations (1d) and (2d)), the nullclines are linear, thus facilitating further analysis (Fig. 3 A-B). The two nullclines (plain lines) delimit three domains in the plane. From wherever of these three domains is the initial starting point, trajectories (Fig. 3B, broken line) return to the unique steady state. This provides robustness towards fluctuations of the receptor concentration.

### 3.2. Influence of parameters values
To assess the influence of the variations of parameters $K_1$, $K_2$ and $K_3$, we studied how the phase plane was modified by varying each of the parameters separately (Smolen et al., 2001; Morohashi et al., 2002; Sriram and Gopinathan, 2005). Three different sets of values for $K_1$, $K_2$ and $K_3$ are explored in Figure 3, giving different pairs of nullclines. $K_2$ range has been explored over two decades (Fig 3C) and only affects the concavity of the nullclines without modifying the qualitative dynamics. Then, keeping the ratio of $K_2$ and $K_3$ constant, we explored two decades of $K_3$ variations and this modifies the area between the two nullclines (Fig 3D). This does not change qualitatively the dynamics of the system. $K_1$ variations were also studied. For $K_1$ greater than 1, the two nullclines cross each other at the stable point that corresponds to an equilibrium between the reactants (Fig 3E) and does not alter the qualitative dynamics otherwise. This qualitative behavior looks like the one of the model with degradation (see below and Fig 5). For $K_2$ lesser than 1 (Fig 3F), the two nullclines cross each other outside of the reachable concentration area but still near the bottom right corner of the phase plane. As in all cases the stable point is kept in the same area, the trajectories are not qualitatively different and the dynamic behavior appears to be robust.

*3.3. Functional interpretation as a 'derivative filter'*

The 'chicken and egg' paradox can be readily explained by this feedback loop model. In cells expressing a mutated Gal4p A699S (where an alanine replaces the serine 699) that cannot be phosphorylated, *GAL* gene activation can initiate but cannot be maintained (Yano and Fukasawa, 1997). Figure 4A shows that in response to the addition of galactose (dashed line), free Gal4p (bold line) increases and initiates transcription of the *GAL3/80* genes. As a consequence, Gal4p binds to new Gal3/80p to produce receptors (plain line) and its free concentration decreases, hence Gal4p cannot maintain transcriptional activity. During the early response phase modelled here, this system acts as a derivative filter (Lauffenburger, 2000; Basu et al., 2004), where galactose is the input step signal and Gal4p is the output peak signal, an approximate derivative of the step input. Furthermore, the model is consistent with the observation that transcription is permanently activated in *gal80Δ* mutant cells (Ideker et al., 2001). Indeed, within the present model, the absence of Gal80p breaks the loop; Gal80p can no longer sequester Gal4p which remains permanently active.

*3.4. Influence of time delays*

As transcriptional interactions generally occur at longer time scales than protein-protein interactions, we experimented with the introduction of time delays (de Jong, 2002) in the differential equations. We introduced a delay in the transcription reaction that led to the equations (1e) and (2e).

$$\frac{d(R)}{dt} = (K_2.S_0 + 1).R_0 - K_2.S_0.R - R(t-\tau) - K_2.R_0.Gal \quad (1e)$$

$$\frac{d(Gal)}{dt} = K_2.S_0.R_0 - K_2.S_0.R - K_2.R_0.Gal \quad (2e)$$

In the simulations shown on figure 4A-C, ten units of galactose are initially introduced at time t=0 in the presence of ten units of receptor. These three simulations correspond respectively to no, short, or long biosynthetic delay. As protein concentrations shift away from equilibrium during the biosynthetic steps, potential chemical energy is stored in the form of Gal3/80p undergoing synthesis. In the absence of delay, this energy is immediately released by the consumption of Gal4p and of the remaining galactose. As the delay increases, more Gal3/80p are being synthesized simultaneously, and therefore more potential energy will suddenly be released at the end of the delay. Thus, the differential time scales of fast protein interactions and slower biosynthetic processes are responsible for the observed feedback acceleration and output signal sharpening.

*3.5. Adaptation of receptor concentration*

To investigate how the system adapts to various galactose inputs, galactose was initially introduced in larger quantity units (15 units) than the receptor (10 units) (Fig. 4D). Compared with an initial dose of ten units of galactose in an otherwise identical experiment (Fig. 4C), a three-step response is observed on figure 4D. Firstly, the receptor is entirely consumed and excess galactose remains as a plateau. Secondly, newly formed receptors rapidly bind the remaining galactose. Thirdly, the initial receptor and Gal4p concentrations are restored. Thus, the receptor concentration is maintained at a constant level, independent of the initial galactose concentration. The *GAL* genes are expressed in proportion to the cellular requirement, at each increase of galactose concentration.

*3.6. Effect of protein degradation*

To investigate the effect of protein degradation, Gal3/80p half-life was set to 1 arbitrary time unit in the presence of a long transcriptional delay of 2 units. It appears that Gal3/80p degradation elicits dampened oscillations of the receptor, Gal4p and Gal3/80p concentrations (Fig. 4E). These oscillations depend on the presence of a transcriptional delay, as shown on figure 4F where Gal4p concentration is monitored in the absence (dashed bold line) or in the presence (bold line) of a long transcriptional delay. As was already shown on panels 3A and 3C, introducing a delay sharpens the time evolution of Gal4p. However, Gal3/80p degradation makes the signal more realistic in that, for instance, Gal4p becomes in excess to Gal3/80p and never reaches zero values.

When Gal3p and Gal80p are distinguished according to the more complete model shown on figure 2A, while including protein degradation, the outcome is similar (Appendix A).

**4. Discussion**

Earlier attempts to model the galactose pathway did not emphasize the feedback loop involving Gal4p activation and induction of Gal3p and Gal80p synthesis. Here we propose that the crucial mechanism generating the ill-understood behavior of the system prior to Gal4p phosphorylation is this feedback loop. A similar approach was recently developed independently, that emphasizes the importance of autoregulation (Ruhela et al., 2004), but with no discussion of the underlying mechanism.

To emphasize the role that feedback may play in the GAL pathway, we simplified our model of the molecular machinery by lumping together Gal3p and Gal80p to the point where its behavior could be described by two differential equations. This simplification has been justified but *aposteriori* it can be also verified. We have shown that the simple model and simulations of the more detailed one have the same qualitative behaviors. In both case, the system is adaptive and acts as a derivative filter (Fig. 3-4 & Fig. 5-6). Degradation modifies moderately the dynamic preventing concentrations to go back down to zero but led to an equilibrium (Fig. 4 E-F & Fig. 5). Both models are robust to parameter variations (Fig. 3 & Fig. 6). The simple two-equation model has several advantages.

Firstly, it captures the key qualitative features of the system dynamics. For instance, phase plane analysis provides useful insight into the poorly understood mechanism of adaptation and high sensitivity to galactose fluctuations. A consequence of the existence of a feedback loop is that receptor concentration is maintained at a constant level, independent of the amount of galactose captured by the cell. Hence, the receptor cannot be saturated by galactose and cells remain sensitive to galactose variations, not to the absolute galactose concentration, without requiring a high number of receptors. Another advantage is that Gal4p-regulated genes are expressed in proportion to cellular requirements, *i.e.* the galactose flux or the time derivative of the galactose concentration.

Secondly, predictions obtained with this model are consistent with previous biological investigations, which were mostly based on the study of one branch of the loop, signal transduction from galactose to Gal4p (Ideker et al., 2001).

Thirdly, the model is rich enough to allow predictions that could be tested at the bench. For instance, consider a mutant strain expressing a Gal4p A699S version to suppress the onlocking of the GAL activation, and Gal2p under the control of a constitutive promoter to avoid permease-dependence of the response to galactose. In these mutant cells, each successive galactose step increase should result in an increase of a Gal1p-GFP chimeric reporter. Furthermore, if the reporter protein is destabilized, its increase should be followed by a decrease that would constitute the signature of the feed-back loop.

The question arises whether the feedback network studied in this paper has some generality beyond the galactose case. Indeed, topological studies (Shen-Orr et al., 2002) on mixed networks that include protein-protein and transcriptional interactions (Yeger-Lotem and Margalit, 2003; Yeger-Lotem et al., 2004) point to the existence of other known cases. A well-known example is the Dig1-Ste12p feedback loop (Bardwell et al., 1998), and longer loops have also been detected, such as Met28-Met4-Cbf1, Mot2-Set1-Ccr4, Ngg1-Ada2-Rtg3-A1-Spt7 or a 13-size feedback loop Sin3-Adh2-Ccr4-Pri2-Swi6-Cdc6-Sfp1-Tec1-Ste12-Tup1-Sps1-Ime1-Rgr1 (Smidtas et al., unpublished). Since interactions may be missing in some of these loops, the interpretation of their roles often requires prior knowledge.

In the future, the model could be refined by considering the relationship between the short-term adaptation described here and the long-term response achieved through Gal4p phosphorylation, or by immersing this system into the wider web of other sugar regulatory pathways and of other types of interaction.

**Acknowledgements**

We are really grateful to Anastasia Yartseva for helpful comments and discussion and K. Sriram and S. Bottani for critically reading this paper. This work was supported by funding from CNRS and Genopole®.

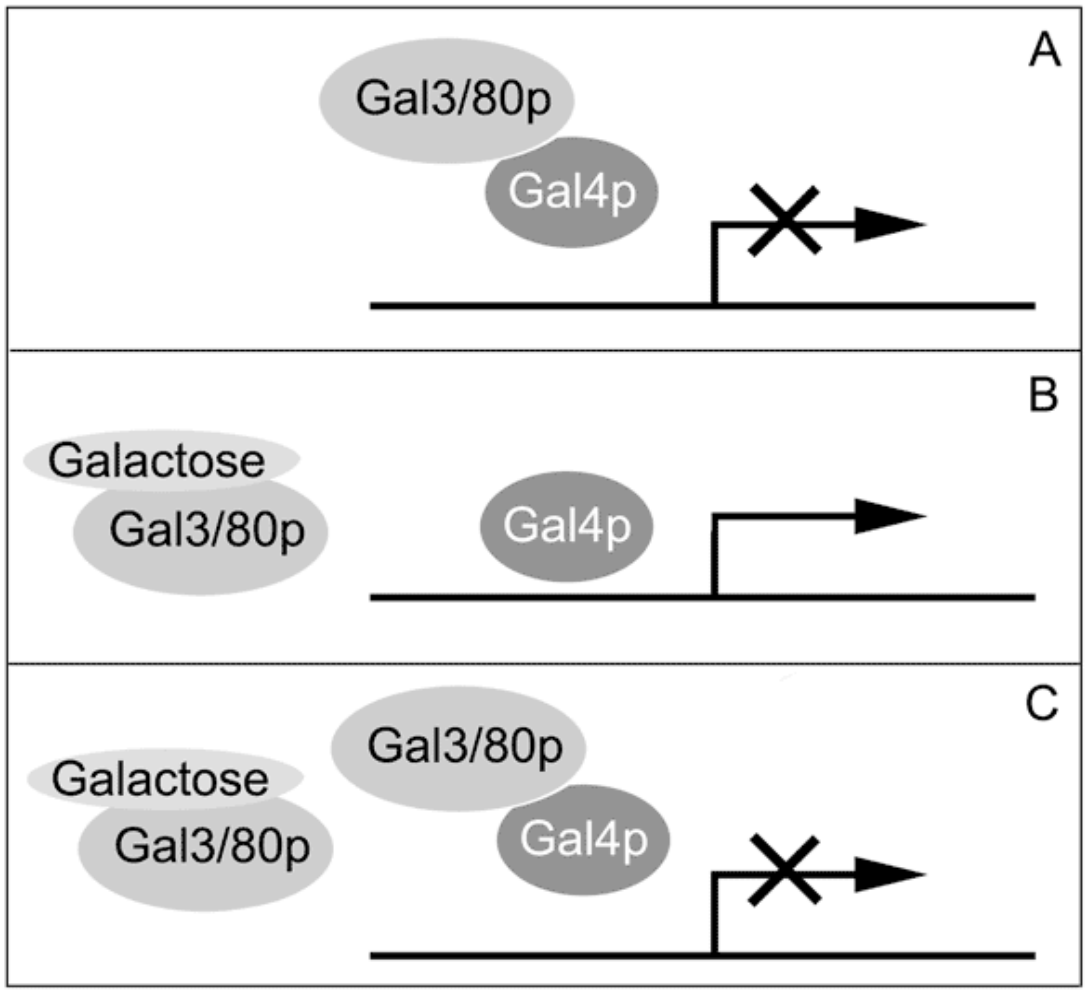

Fig. 1. Diagrammatic representation of the galactose induction loop.
(A) In the absence of galactose, the transcriptional activity of Gal4p is inhibited by Gal3/80p. (B) The association of galactose with Gal3/80p allows Gal4p to be freed from Gal80p inhibition and to activate transcription of new Gal3/80p. (C) Newly synthesized Gal3/80p inhibits the transcriptional activity of Gal4p.

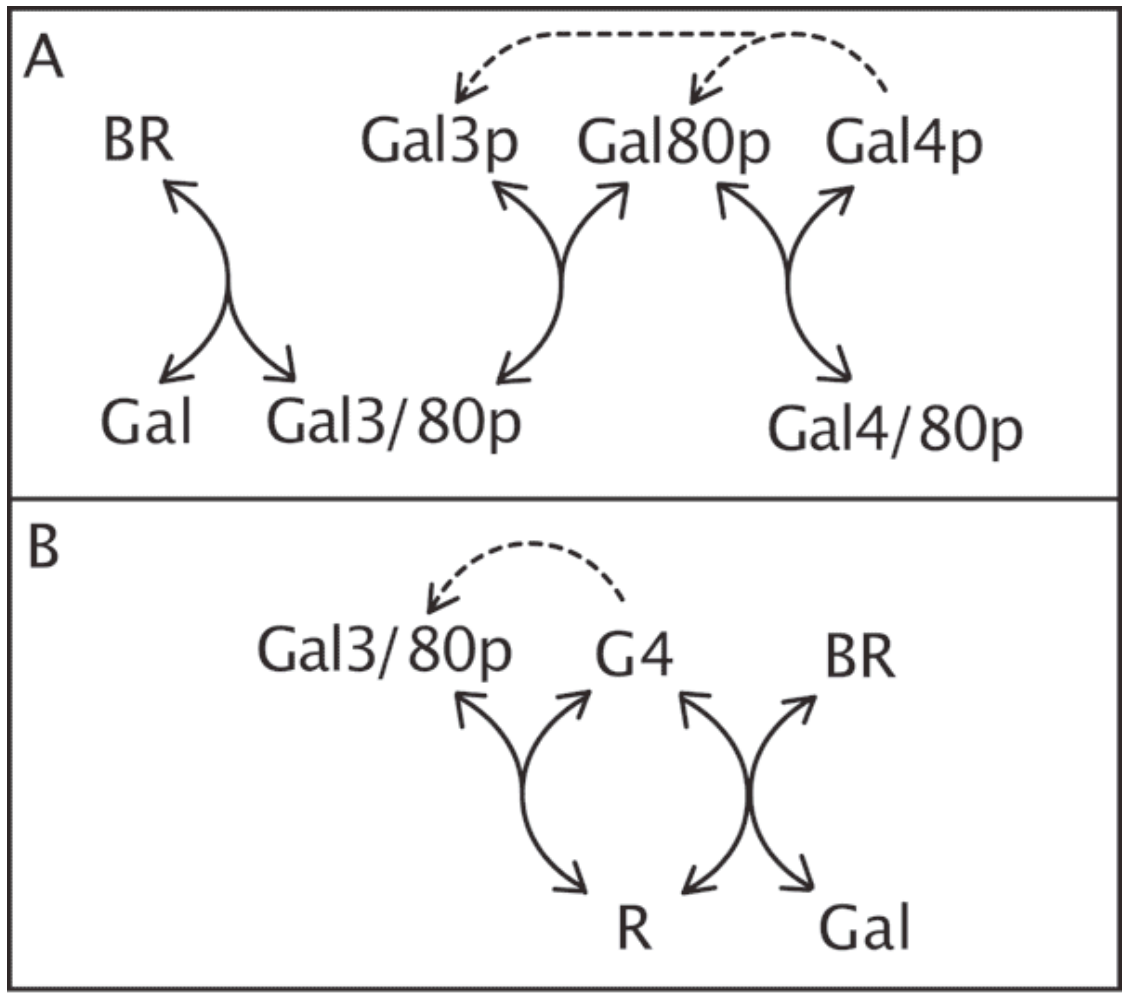

Fig. 2. GAL core regulatory pathway.
(A) Detailed model. In the presence of galactose, Gal3p, Gal80p and galactose ('Gal') bind together, thus decreasing the binding of Gal80p to Gal4p. Without Gal80p, Gal4p becomes active and induces transcription (dashed arrows) of the *GAL3* and *GAL80* genes. This closes the feedback loop, as newly synthesized Gal3p and Gal80p shift the equilibrium back towards Gal4p inactivation (Gal4p/80p). (B) Simplified model. Receptor ('R') denotes Gal4/3/80p; Bound Receptor ('BR') denotes the Gal3/80p/galactose complex. In the absence of galactose, Gal3/80p sequesters Gal4p ('G4') into the Receptor, thus preventing its transcriptional activity. In the presence of galactose, Gal4p is released and induces transcription of the *GAL3* and *GAL80* genes. Newly synthesized Gal3/80p shift the equilibrium back towards Gal4p inactivation.

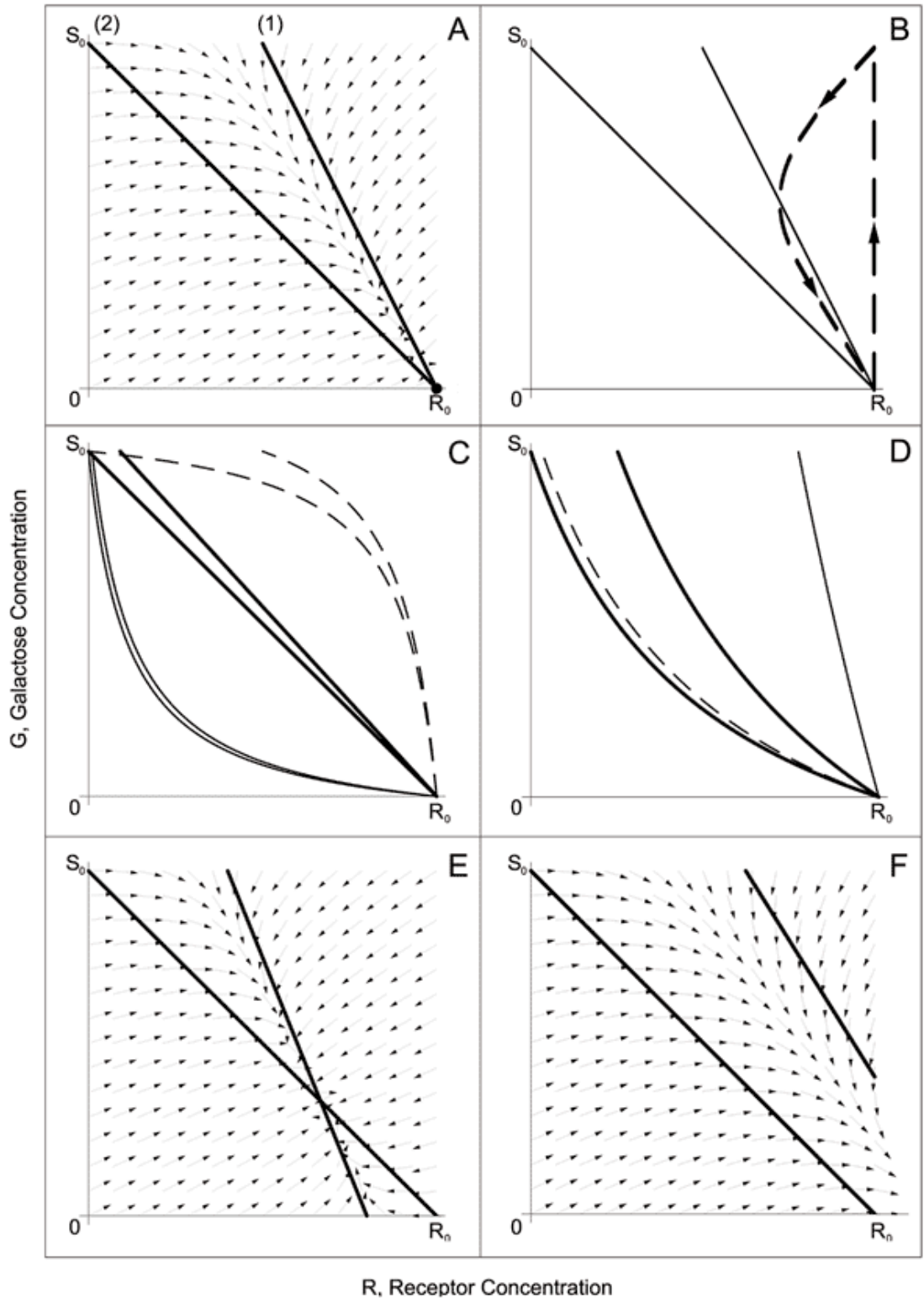

Fig. 3. Phase portrait and parameter variation study of the two-equation system.

The curves labeled 1 or 2 are nullclines that correspond to the situations where: 1) receptor synthesis and consumption by galactose fixation are balanced, or 2) free galactose addition and galactose fixation are balanced. There is a single stable point where the two curves cross each other. (A-B) Phase portrait for equations (1d) and (2d). Simplified phase portrait and trajectory for equations (1d) and (2d). The nullclines (plain lines) are represented for $K_1=1$ and $K_2=K_3$. Arrow pairs represent the slopes in each of the three domains defined by the two nullclines. (B) A typical trajectory is represented by a broken line. In the absence of galactose, the system is at the stable point in the lower right corner. When galactose is introduced, all galactose is in its free form and the state moves to the upper right corner. Then, galactose binds to the receptor and causes free galactose and receptor concentrations to decrease. This binding causes Gal4p to activate transcription, thereby producing new receptor whose concentration increases back to the stable point. (C-F) Phase portrait for equations (1c) and (2c). These curve pairs are represented for various values of $K_1$, $K_2$ and $K_3$. (C) $K_1=1$, $K_2=10$, $K_3=10$ (bold line), $K_1=1$, $K_2=100$, $K_3=10$ (plain line), $K_1=1$, $K_2=1$, $K_3=10$ (dashed line). (D) $K_1=1$, $K_2=3$, $K_3=1$ (bold line), $K_1=1$, $K_2=30$, $K_3=10$ (plain line), $K_1=1$, $K_2=0.3$, $K_3=0.1$ (dashed line). (E) $K_1=1.5$, $K_2=1$, $K_3=1$. (F) $K_1=0.6$, $K_2=1$, $K_3=1$.

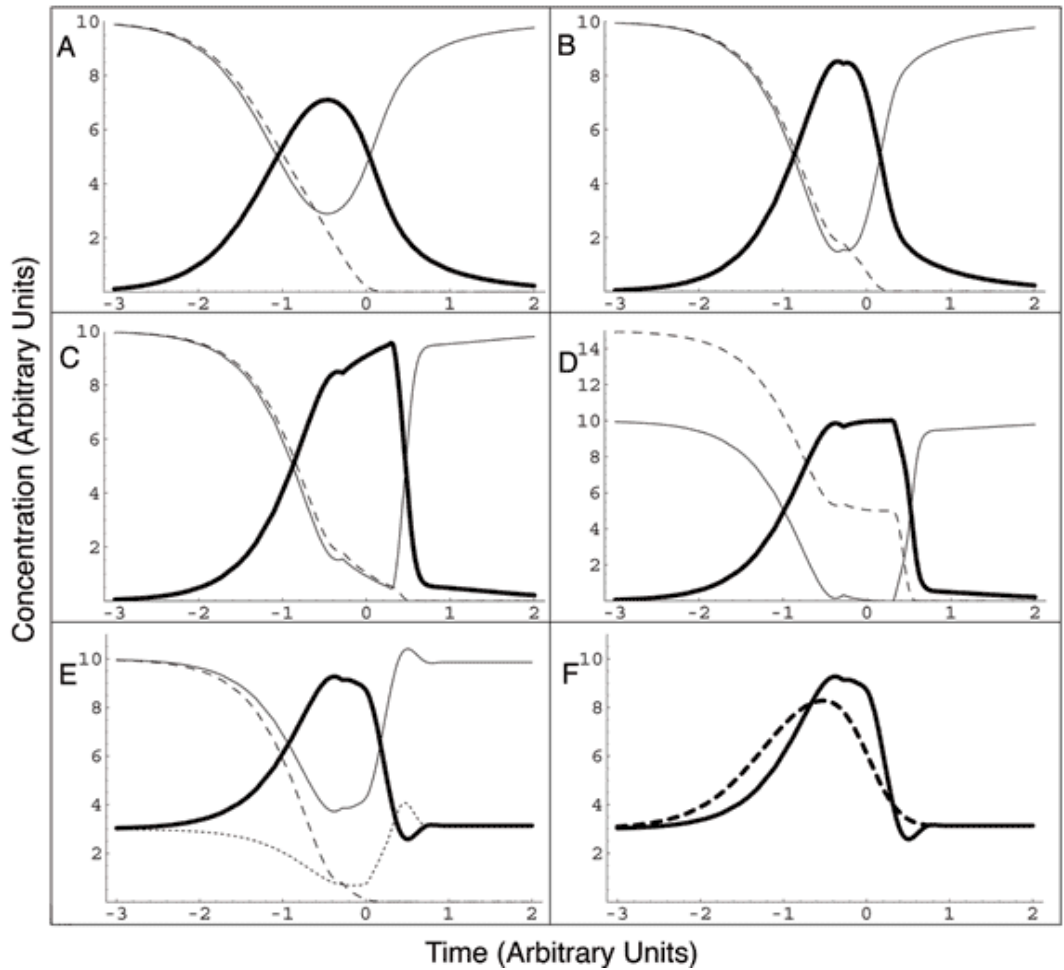

Fig. 4. Simulations of the GAL System time response depicted in equations (1-3).

The initial state consists of ten units of receptor (plain line), no free Gal3/80p away from the receptor, and no free Gal4p/DNA (bold line) unless otherwise stated. Ten units of galactose (dashed line) are initially introduced, unless otherwise stated. Note the log scale on the time axis. The Mathematica function NDelayDSolve written by Allan Hayes was used for computation. Curves A-E show computed time responses for different sets of parameters, as follows: (A) No delay in transformations. (B) Short transcriptional delay of 0.5 arbitrary units of time (a.u.). (C) Long transcriptional delay of 2 a.u. (D) Long transcriptional delay. Fifteen units of galactose are initially introduced, and the concentration scale has been changed to reflect this change. (E) Long transcriptional delay, and Gal3/80p (dotted line) degradation (half-life of 1 a.u.). Although no additional free Gal4p has been introduced, Gal3/80p decay results in excess Gal4p that is not sequestered in the receptor, hence the non-zero value of initial and final Gal4p. (F) Long transcriptional delay (bold line), compared to the absence of delay (dashed line), and Gal3/80p degradation as in (E). Only Gal4p is represented here (same remark as in E).

**Appendix A**

The detailed model includes the following equations corresponding to Figure 2A. This model includes slow Gal3/80p degradation (Ruhela et al., 2004) and distinguishes Gal3p and Gal80p. Altogether, it involves more parameters than the reduced model of the main text. Several of these additional parameters do not help to analyze the feedback loop which is the focus of this study. Yet they allow for a more realistic simulation. It is possible to introduce time delays corresponding to the biosynthetic reactions.

$$c \times \text{Gal4p/DNA} \xrightarrow{k_1} c \times \text{Gal4p/DNA} + \text{Gal3p} + \text{Gal80p}$$

$$\text{Gal80p} + \text{Gal4p/DNA} \underset{k_3}{\overset{k_2}{\rightleftarrows}} \text{Gal80p/Gal4p/DNA}$$

$$\text{Gal3p} + \text{Gal80p} \underset{k_5}{\overset{k_4}{\rightleftarrows}} \text{Gal3p/Gal80p}$$

$$\text{Gal3p/Gal80p} + \text{Galactose} \xrightarrow{k_6} \text{Galactose/Gal3p/Gal80p}$$

$$\text{Gal3p} \xrightarrow{k_7} \text{degradation}$$

$$\text{Gal80p} \xrightarrow{k_7} \text{degradation}$$

The model has been simulated in Mathematica (Fig. 5) with the following equations and parameters.

$d(\text{Gal4p|DNA})/dt = -k_2.\text{Gal4p|DNA}.\text{Gal80p} + k_3.\text{Gal80p|Gal4p|DNA}$

$d(\text{Gal3p})/dt = k_1.\text{Gal4p|DNA}^c - k_4.\text{Gal3p}.\text{Gal80p} + k_5.\text{Gal3p|Gal80p} - k_7.\text{Gal3p}$

$d(\text{Gal80p})/dt = k_1.\text{Gal4p|DNA}^c - k_4.\text{Gal3p}.\text{Gal80p} + k_5.\text{Gal3p|Gal80p} - k_7.\text{Gal80p}$
$\qquad -k_2.\text{Gal4p|DNA}.\text{Gal80p} + k_3.\text{Gal80p|Gal4p|DNA}$

$d(\text{Gal3p|Gal80p})/dt = -k_5.\text{Gal3p|Gal80p} + k_4.\text{Gal3p}.\text{Gal80p}$
$\qquad -k_6.\text{Gal3p|Gal80p}.\text{Galactose} + k_7.\text{Galactose|Gal3p|Gal80p}$

$d(\text{Gal80p|Gal4p|DNA})/dt = k_2.\text{Gal80p}.\text{Gal4p|DNA} - k_3.\text{Gal80p|Gal4p|DNA}$

$d(\text{Galactose})/dt = -k_6.\text{Gal3p|Gal80p}.\text{Galactose}$

The various parameters of the complete dynamic model used in the simulations are:
$k_1=1$, $k_2=1$, $k_3=1$, $k_4=1$, $k_5=1$, $k_6=1$, $k_7=1E-4$, $c=4$
Initial concentrations are: Gal4p/DNA(0)=0.1, Gal3p(0)=1, Gal80(0)=1, Gal3p/Gal80p(0)=1, Gal80p/Gal4p/DNA(0)=0.1, Galactose(0)=2

*Influence of parameter values in the previous model*

Biochemical parameters are expected to vary somewhat from cell to cell and from one member of a species to another. Furthermore, given the uncertainty that undoubtedly exists for parameters, we need to test this model with parameter variation. To assess how the qualitative behavior of this model is robust to parameter variations, the main characteristic features of the Gal4p response were observed. The two-step response, increase and decrease in Gal4p that implement adaptation (Lauffenburger, 2000), should be robust to variations in parameters. Figure 6 plots the time $t_m$ at which the peak of Gal4p response concentration is reached, the amplitude $a_m$ of this response and the decrease $d_m$ from the peak to t=4. These three values are illustrated on Figure 5. Results of the parameter variations exploring two orders of magnitude for seven parameters $k_1$-$k_7$ and ±50% variation for c are shown on Figure 6. The main qualitative features, perfect or partial adaptation, of the Gal4p signal is preserved in all simulations and the response appearance never varied dramatically from the control with no parameter change. The average amplitude (respectively time) is 0.8 (resp. 2), standard deviation of the amplitude (resp. time) is 0.1 (resp. 0.4). Dynamically, there is no qualitative difference compared to the simplified model (Fig. 4E-F) when it also includes degradation, even with a wide range of parameter variations.

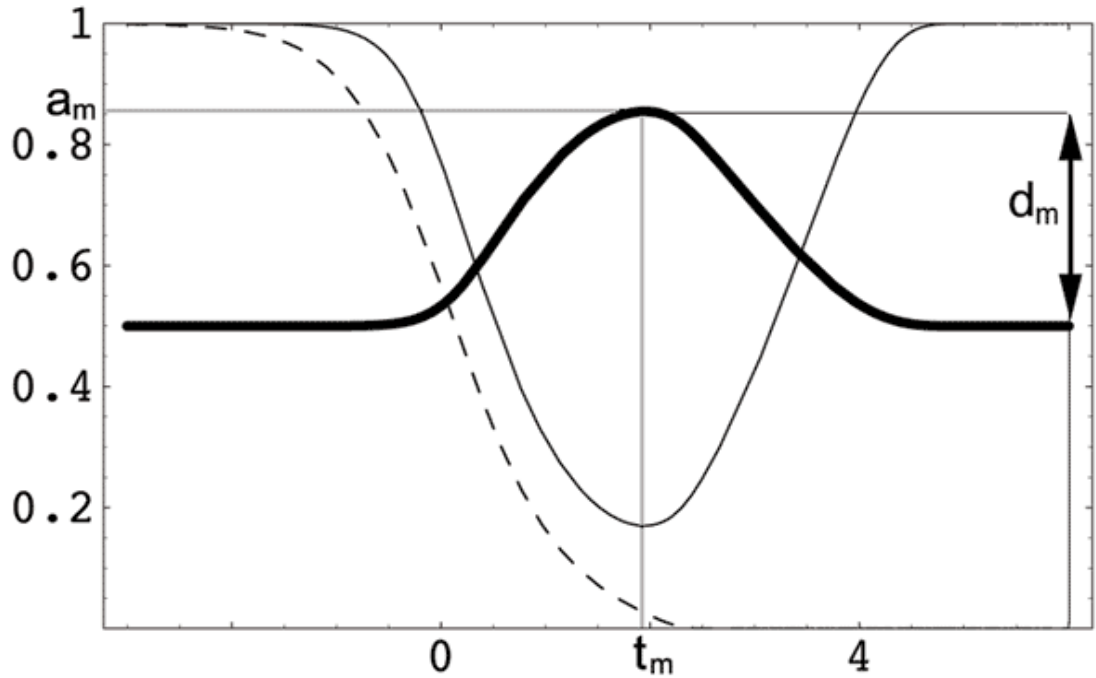

Fig. 5. Time response of the GAL pathway depicted in figure 2A. The initial state consists of 0.1 unit of Gal4p/DNA (bold line), and 1 unit of Gal80p (plain line). Two units of galactose (dashed line) are initially introduced, unless otherwise stated. Note the log scale on the time axis. Along the ordinate axis, the Gal4p scale has been amplified five-fold, and the galactose scale has been decreased two-fold. This simulation involves no delay in transformations. In this respect, the result may be compared with Figure 4A.

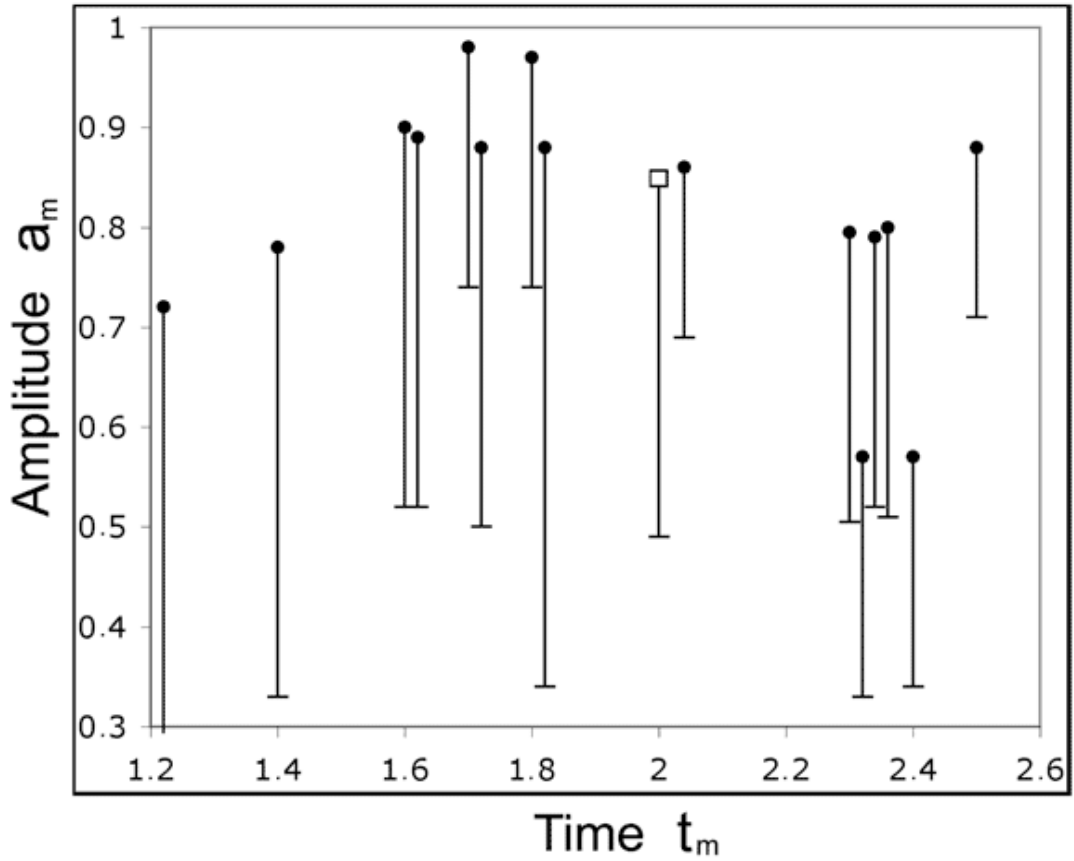

Fig. 6. Robustness of the non-reduced model to parameter variations. The scatter plot displays the Amplitude $a_m$ versus Time $t_m$ of the maximal signal response of Gal4p (as defined in Fig.5) for 10-fold increase or decrease of parameter k1-k7 and ±50% variation in parameter c. The square represents the model with no parameter variation. Each vertical bar indicates how much the Gal4p response decreases after the maximal signal response until time t=6 ($d_m$ distance in Figure 5).